\documentclass[journal=nalefd,manuscript=article]{achemso}

\usepackage[version=3]{mhchem} 
\usepackage{amsmath}
\usepackage[numbers,sort&compress]{natbib}
\usepackage{graphicx}
\usepackage{textcomp}
\usepackage{braket}
\setkeys{acs}{maxauthors=99, etalmode=truncate}

\author{Jianpei Geng}
\affiliation{School of Physics, Hefei University of Technology, Hefei 230009, China}
\altaffiliation{These authors contributed equally to this work.}
\email{jianpei.geng@hfut.edu.cn}

\author{Xuankai Zhou}
\affiliation{3rd Institute of Physics, University of Stuttgart, Stuttgart, 70569, Germany}
 \altaffiliation{These authors contributed equally to this work.}

\author{Nils Gross}
\affiliation{Max Planck Institute for Solid State Research, Stuttgart, 70569, Germany}

\author{Song Li}
\affiliation{HUN-REN Wigner Research Centre for Physics, P.O. Box 49, H-1525 Budapest, Hungary}

\author{Yan Tung Kong}
\affiliation{3rd Institute of Physics, University of Stuttgart, Stuttgart, 70569, Germany}

\author{Jixing Zhang}
\affiliation{3rd Institute of Physics, University of Stuttgart, Stuttgart, 70569, Germany}

\author{Guodong Bian}
\affiliation{HUN-REN Wigner Research Centre for Physics, P.O. Box 49, H-1525 Budapest, Hungary}

\author{San Lam Ng}
\affiliation{3rd Institute of Physics, University of Stuttgart, Stuttgart, 70569, Germany}

\author{Cheng-I Ho}
\affiliation{3rd Institute of Physics, University of Stuttgart, Stuttgart, 70569, Germany}

\author{Andrej Denisenko}
\affiliation{3rd Institute of Physics, University of Stuttgart, Stuttgart, 70569, Germany}

 \author{Rainer Stöhr}
\affiliation{3rd Institute of Physics, University of Stuttgart, Stuttgart, 70569, Germany}

\author{Ruoming Peng}
\affiliation{3rd Institute of Physics, University of Stuttgart, Stuttgart, 70569, Germany}
\email{ruoming.peng@pi3.uni-stuttgart.de}

\author{Jurgen Smet}
\affiliation{Max Planck Institute for Solid State Research, Stuttgart, 70569, Germany}

\author{Jörg Wrachtrup}
\affiliation{3rd Institute of Physics, University of Stuttgart, Stuttgart, 70569, Germany}
\alsoaffiliation{Max Planck Institute for Solid State Research, Stuttgart, 70569, Germany}

\title[An \textsf{achemso} demo]
  {Deformation-Driven Enhancement of Spin Defect Emission in Hexagonal Boron Nitride}

\abbreviations{IR, NMR, UV}

\begin{document}


\begin{abstract}
  The negatively charged boron vacancy ($V_\text{B}^-$) in hexagonal boron nitride (hBN) has been extensively investigated as it offers a novel playground for two-dimensional quantum sensing, with ultimate proximity to target samples. However, its practical sensitivity is limited by the intrinsically weak photoluminescence of the spin ensemble. Here, we report a photoluminescence enhancement of up to 30 times from $V_\text{B}^-$ centers in suspended regions of hBN compared to those in substrate-supported areas. The key spin properties, such as the optically detected magnetic resonance (ODMR) contrast and linewidth, as well as the spin lifetime, of the $V_\text{B}^-$ centers in this region are well preserved. Detailed investigations, including measurements of zero-field ODMR, Raman spectroscopy, and Kelvin probe force microscopy, reveal a correlation between emission enhancement and local deformation in the sample. It is concluded that the suspended regions exhibit higher local deformation compared to the supported areas, breaking the local symmetry and thereby activating otherwise forbidden or weak optical transitions of the $V_\text{B}^-$ centers.
\end{abstract}

\section{Introduction}

Solid-state spin defects\cite{wolfowicz2021quantum} hold great promise for advancing modern quantum sensing technologies\cite{degen2017quantum}. Spin defects in bulk materials, such as nitrogen-vacancy (NV) centers in diamond\cite{jelezko2006single,doherty2013nitrogen} and silicon vacancies in silicon carbide\cite{koehl2011room}, offer exceptional sensitivity and a broad dynamic range\cite{barry2020sensitivity,du2024single}. However, bringing these spin defects close to the surface often degrades their spin properties, posing a significant challenge in surface termination. Recently, hexagonal boron nitride (hBN)\cite{cassabois2016hexagonal,caldwell2019photonics} has garnered growing attention as a promising alternative for surface sensing. Its intrinsic two-dimensional (2D) nature allows hBN to host spin defects down to the monolayer limit, enabling exceptional proximity to the target while potentially mitigating surface-induced decoherence common in bulk materials. As a 2D insulator, hBN supports a variety of optically addressable spin defects\cite{gottscholl2020initialization,gottscholl2021room,chejanovsky2021single,stern2022room,guo2023coherent,stern2024quantum}, among which the negatively charged boron vacancy ($V_\text{B}^-$) has emerged as a promising candidate for various sensing applications\cite{gottscholl2021spin}. Sharing spin configurations similar to those of NV centers in diamond\cite{gottscholl2020initialization}, $V_\text{B}^-$ centers in hBN have been successfully employed to probe magnetic responses in 2D magnets\cite{healey2023quantum,huang2022wide}, visualize magnon propagation\cite{zhou2024sensing}, and detect paramagnetic spins in solution\cite{gao2023quantum,robertson2023detection}.

Many efforts have been made to optimize the performance of $V_\text{B}^-$ centers, including advancements in implantation techniques\cite{zabelotsky2023creation,sarkar2023identifying,hennessey2024framework}, isotope purification\cite{haykal2022decoherence,clua2023isotopic,janzen2024boron,gong2024isotope}, spin control\cite{liu2022coherent,gao2022nuclear,rizzato2023extending,gong2023coherent,ramsay2023coherence}, and other optimization processes\cite{zhou2023dc,whitefield2023magnetic,gale2023manipulating}. Despite these efforts, a major challenge remains within their optical characteristics: the radiative emission of $V_\text{B}^-$ centers is rather weak, and only limited fluorescence counts can be collected from a $V_\text{B}^-$ ensemble\cite{reimers2020photoluminescence,ivady2020ab}. To enhance the fluorescence emission, earlier attempts have focused on engineering the photonic environment. A Purcell enhancement of the radiative decay has been demonstrated by coupling the $V_\text{B}^-$ centers to photonic or plasmonic resonators\cite{gao2021high,mendelson2022coupling,xu2022greatly}. However, even with this enhancement, the emission is still dim, resulting in a bad signal-to-noise ratio, thereby limiting potential applications. Given the appealing perspectives on quantum sensing technologies, exploring more methods for further enhancement of the $V_\text{B}^-$ fluorescence is still important to achieve optimal performance.

In this report, we present an alternative approach to enhance the radiative emission of the $V_\text{B}^-$ ensemble in hBN. We achieve an up to 30-fold increase in brightness for $V_\text{B}^-$ centers in a suspended hBN region compared to supported regions. Notably, all the fluorescence enhancement can directly contribute to the improved sensitivity, as key spin properties, such as the optically detected magnetic resonance (ODMR) contrast and linewidth, as well as the spin lifetime, are well preserved. Through detailed investigations using zero-field ODMR, Raman spectroscopy, and Kelvin Probe Force Microscopy (KPFM), we have identified a direct spatial correlation between the enhanced emission and the deformation potential of the local hBN lattice. The deformation is accompanied by a reduction of the symmetry, which could enhance the dipole transitions of the $V_\text{B}^-$ centers\cite{ivady2020ab} and lead to stronger fluorescence emission. Our results coincide with a previous observation that color centers at creases in hBN flakes appear to be brighter\cite{curieCorrelativeNanoscaleImaging2022}. Based on these findings, we suggest that local-deformation induced symmetry breaking could serve as a powerful mechanism to enhance the emission of dim spin defects such as $V_\text{B}^-$ centers.

\begin{figure}[h!]
  \centering
  \includegraphics[width = \textwidth]{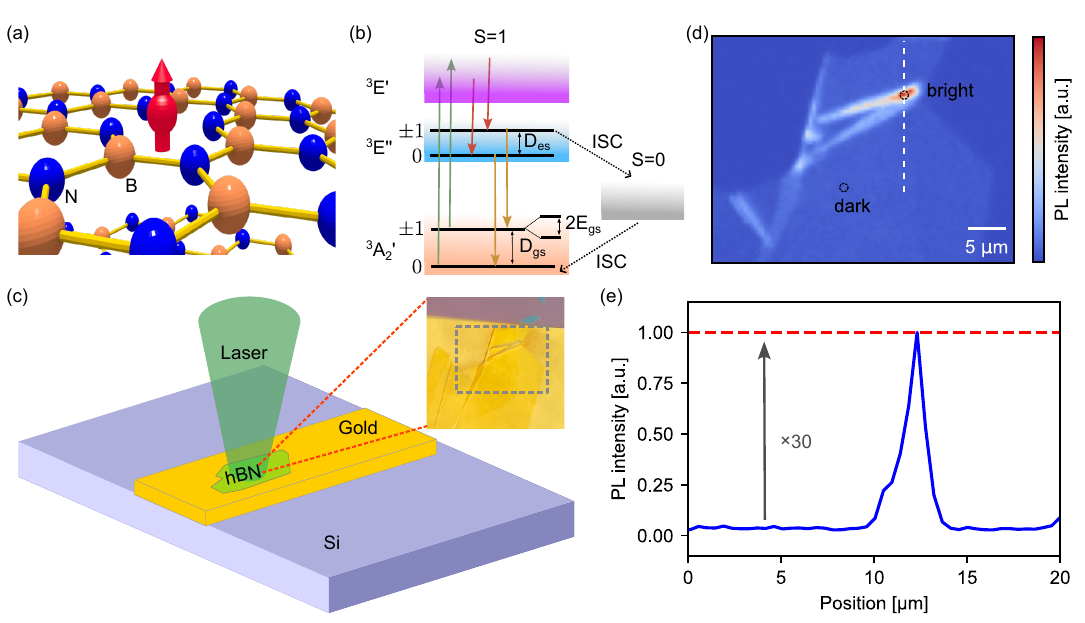}
  \caption{(a) Atomic structure of the ${V_\text{B}^-}$ center in an hBN crystal, with the electron spin of the ${V_\text{B}^-}$ center highlighted in red.
  (b) The electronic energy levels of the ${V_\text{B}^-}$ center. The lower energy excited state is shaded in blue, and evolves through the intersystem crossing (ISC) process. 
  (c) A schematic of the device of the ${V_\text{B}^-}$ centers and an optical image of the hBN sample S1. The hBN flake is transferred onto a gold stripline.
  (d) A 2D confocal scan of the hBN sample S1 in the region enclosed by the dashed rectangle in (c), showing bright and dark spots as marked with dashed circles on the map.
  (e) A linecut of the 2D confocal scan along the white dashed line in (d). 
  }
  \label{fig1}
\end{figure}

\section{Enhanced fluorescence emission of the ${V_\text{B}^-}$ centers}
The ${V_\text{B}^-}$ center is a negatively charged spin defect manifested by a missing boron atom surrounded by 3 equivalent nitrogen atoms in the hBN lattice, as illustrated in Fig. \ref{fig1}a. Therefore, the defect has a D$_\text{3h}$ point group symmetry. Fig. \ref{fig1}b shows the schematic energy-level structure of the ${V_\text{B}^-}$ center. The ground state $^3A_2'$, the excited states $^3E''$ and $^3E'$, and at least one metastable state are involved in the photodynamics to explain the fluorescence emission. The $^3A_2'$, $^3E''$, and $^3E'$ are spin-triplet states. Owing to the D$_\textrm{3h}$ symmetry, the optical transition between the ground state $^3A_2'$ and the excited state $^3E''$ is in first order symmetry-forbidden \cite{ivady2020ab}. Consequently, the dipole-allowed transition from $^3A_2'$ to $^3E'$ is typically employed to excite ${V_\text{B}^-}$ using a green laser. Once excited to $^3E'$, the system rapidly relaxes to $^3E''$ via internal conversion. The subsequent relaxation from $^3E''$ to $^3A_2'$ proceeds mainly through nonradiative decay and an intersystem crossing involving the metastable state. 
The radiative transition from $^3E''$ to $^3A_2'$ contributes to the fluorescence emission but is intrinsically weak, being symmetry-forbidden in first order and only allowed in second order. 
As a result, the fluorescence emission from the ${V_\text{B}^-}$ ensemble is notably dim. Thus, an emission enhancement is expected by a reduction of the symmetry induced by the local deformation of the hBN lattice. 

We fabricated multiple hBN devices to investigate the optical and spin properties of the ${V_\text{B}^-}$ centers. A schematic of the device is shown in Fig. \ref{fig1}c.
Multilayer hBN flakes were mechanically exfoliated and transferred onto a stripline structure composed of a 10/200 nm Ti/Au layer deposited on the 300 nm SiO$_2$/Si substrate. The shallow ${V_\text{B}^-}$ spin ensembles were subsequently created via He$^{+}$ ion implantation at the energy of 2.5 keV and the dose of $10^{14}$ ${\text{cm}}^{-2}$. 
The stripline was engineered to deliver microwave fields for efficiently driving the spin transitions of the ${V_\text{B}^-}$ centers. 
The inset of Fig. \ref{fig1}c shows an optical image of such a device, labeled as sample S1. We observed wrinkles on the hBN flake, which can form during the transfer process. The area nearby the cross of the two neighbouring wrinkles appeared to be suspended.
After implantation, we mounted sample S1 on a home-built confocal setup and excited the ${V_\text{B}^-}$ centers with a 514 nm green laser. The photoluminescence was collected using a silicon avalanche photodiode (APD). We first measured the confocal map by scanning the laser spot across the hBN flake and collecting the photoluminescence. Fig. \ref{fig1}d shows the confocal map of the region enclosed by the dashed rectangle in the inset of Fig. \ref{fig1}c. The shape of the bright region in the confocal map coincides with the wrinkles in the optical image. The brightest spot appears in the suspended area nearby the cross of the two neighbouring wrinkles. The fluorescence intensity of the brightest spot is more than 30 times larger than that in the pristine flat area, as is shown in Fig. \ref{fig1}e. The ODMR spectra measured later confirm that the emission in both areas comes from the ${V_\text{B}^-}$ centers.

\section{Spin properties of the ${V_\text{B}^-}$ centers}

\begin{figure}[h!]
  \centering
  \includegraphics[width = \textwidth]{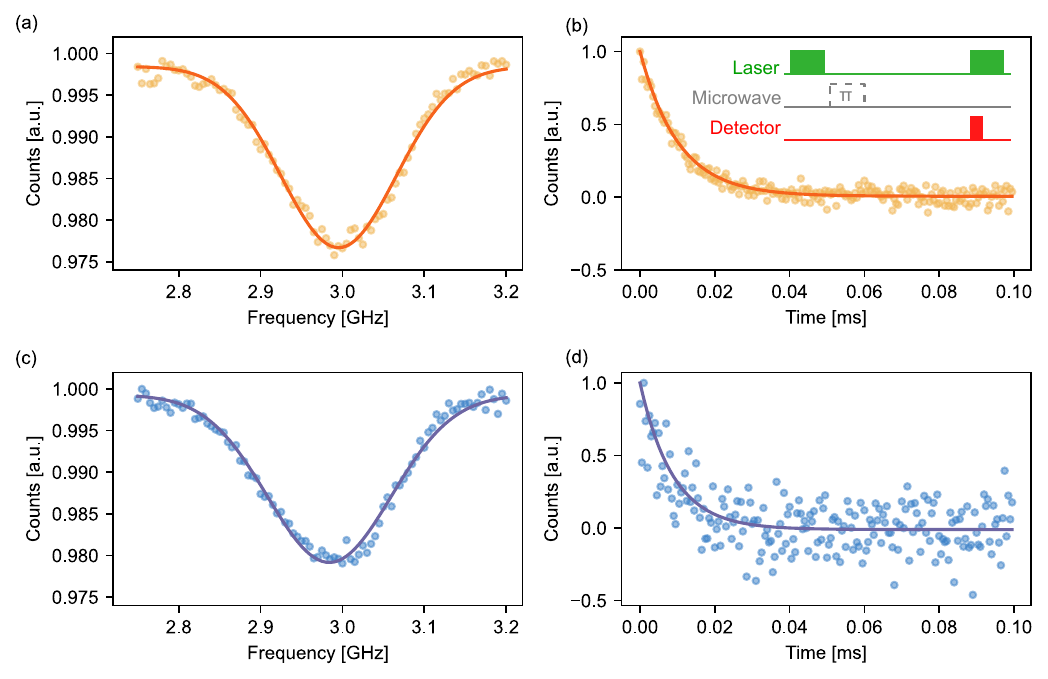}
  \caption{ (a, c) Continuous-wave ODMR spectra of the ${V_\text{B}^-}$ defects in the bright spot (a) and the dark spot (c). The dots show the experimental data and the solid lines are the Gaussian fit. 
  (b, d) Spin relaxation time ($T_1$) measurements for the bright spot (b) and the dark spot (d). The dots show the experimental data and the solid lines are the exponential fit. The measurement sequence is shown in the inset of b.
  }
  \label{fig2}
\end{figure}

To characterize spin properties of the ${V_\text{B}^-}$ centers, we performed continuous-wave (CW) ODMR measurements at spots in both bright and dark hBN regions at a bias out-of-plane magnetic field of about 17 mT. 
Fig. \ref{fig2}a shows the CW ODMR spectrum of the brightest spot which is marked with a dashed circle as a "bright" spot in Fig. \ref{fig1}d. The resonant frequency of about 3.0 GHz coincides with the spin transition between $m_s=0$ and $m_s=-1$ sublevels of the $V_\text{B}^-$ ground state. The fit of the spectrum gives a linewidth of 165(3) MHz and a contrast of 2.18(3)\%. Rabi oscillations further reveal a contrast above 10\% (See Section 8, SI). The large linewidth is attributed to interactions of the $V_\text{B}^-$ electron spin with surrounding nuclear spins. For comparison, Fig. \ref{fig2}c shows the CW ODMR spectrum of a dark spot that is marked with the other dashed circle in Fig. \ref{fig1}d. The spectrum shows the same resonant frequency as that in Fig. \ref{fig2}a within uncertainties, indicating that both spectra correspond to $V_\text{B}^-$ centers. The fit of the spectrum of the dark spot gives a linewidth of 178(4) MHz and a contrast of 2.01(3)\%. The spectrum of the bright spot exhibits a slightly narrower linewidth and similar contrast compared to the dark spot. 

Moreover, we performed spin relaxation ($T_1$) measurements of the $V_\text{B}^-$ centers at both bright and dark spots. To avoid the influence of charge-state dynamics, the spin-relaxation process is measured with an initial state of both $m_s=0$ and $m_s=-1$. The difference of the measurement results is robust to common-mode noise and spin-independent processes such as charge-state dynamics. The measurement pulse sequence is shown in the inset of Fig. \ref{fig2}b. A laser pulse is applied to initialize the $V_\text{B}^-$ centers into the $m_s=0$ spin state. For the measurement with initial state of $m_s=-1$, a microwave $\pi$ pulse is used to flip the spin state. Starting from $m_s=0$ or $m_s=-1$, the spin state of the $V_\text{B}^-$ centers relaxes towards equilibrium during the following time.
A second laser pulse is then applied to read out the state population in $m_s=0$ as a function of the time for relaxation. The difference between the measurement results with initial states of $m_s=0$ and $m_s=-1$ is shown in Fig. \ref{fig2}b for the bright spot and in Fig. \ref{fig2}d for the dark spot, respectively. By fitting we obtain a respective relaxation time of $T_1=10.3(3) \ \mu$s and $T_1=9(1) \ \mu$s for the bright and dark spots. The relaxation time is the same within uncertainties and similar to other reports of He$^+$-implanted hBN samples\cite{gong2023coherent}. The preserved spin relaxation time, together with slightly improved ODMR linewidth and contrast, highlights no degradation of spin properties for the bright spot. Therefore, the fluorescence enhancement could directly lead to an improvement in the sensitivity of $V_\text{B}^-$ centers in the bright spot compared to the dark spot.

The fluorescence intensity of a ${V_\text{B}^-}$ ensemble is dependent on the number of ${V_\text{B}^-}$ centers and the fluorescence emission of each ${V_\text{B}^-}$ center. In the following, we exclude the possibility that the bright spot in the suspended region contains more ${V_\text{B}^-}$ centers than the dark spot in the substrate-supported region. The ${V_\text{B}^-}$ centers were created by He$^{+}$ ion implantation with an homogeneous He$^{+}$ dose over the area of the confocal map. To investigate the influence of the suspended or supported region on the creation of vacancies, we performed SRIM (the Stopping and Range of Ions in Matter) simulations on collisions between the He$^{+}$ ion and atoms in hBN on a substrate (see Supplemental Section 2). The substrate layer was set as gold for the supported region and as N$_2$ gas to approximate the condition of the suspended region. The simulation results show that the vacancy concentration in the case of the N$_2$ substrate is slightly lower and the same for thin and thick hBN layers, respectively, compared to that in the case of the gold substrate. This indicates that the ion implantation process did not create a higher density of ${V_\text{B}^-}$ centers in the suspended region. 

Another possibility would be that the bright spot contained more ${V_\text{B}^-}$ centers because it consisted of both the suspended region and the surrounding boundary and therefore had a larger volume. This possibility could be ruled out by the ODMR measurements shown in Fig. \ref{fig2}a and c. Since the magnetic field was applied perpendicular to the plane of the flat hBN area, it was aligned along the principal symmetry axis of the ${V_\text{B}^-}$ centers in both flat suspended and supported regions, but misaligned at the boundary. Therefore, the ODMR spectrum of the ${V_\text{B}^-}$ centers at the boundary would present a resonant frequency much different from that of the flat suspended or supported region. If the ${V_\text{B}^-}$ centers at the boundary contributed significantly to the bright spot, compared to the dark spot, the ODMR spectrum of the bright spot should show a broader linewidth and lower contrast since it mixed the spectra of the ${V_\text{B}^-}$ centers at the boundary and in the suspended region (see Section 3, SI). This was in contrast to the experimental observations shown in Fig. \ref{fig2}a and c, ruling out the possibility that the ${V_\text{B}^-}$ centers at the boundary contributed to a larger number of ${V_\text{B}^-}$ centers in the bright spot. 
\section{Correlation between enhanced fluorescence and deformation}

\begin{figure}[h!]
  \centering
  \includegraphics[width = \textwidth]{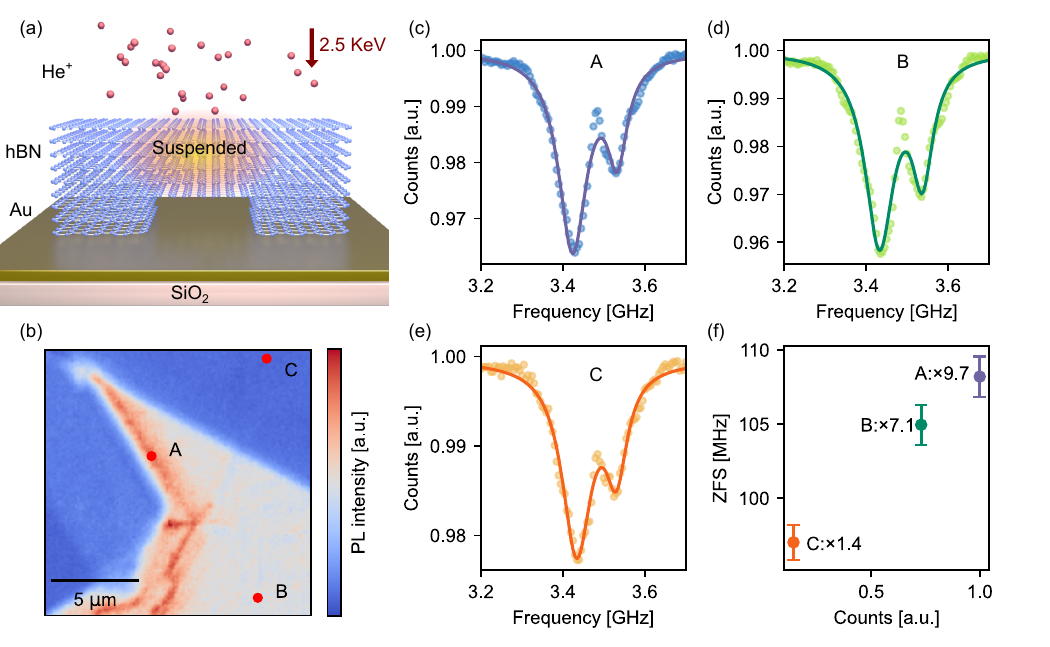}
  \caption{(a) Schematic cross section of a hBN flake with a suspended region during He$^+$ ion implantation. The hBN flake that varies in thickness is placed upside down on the substrate during the transfer process. The thinner region of the hBN is suspended after the transfer process.
  (b) Confocal photoluminescence (PL) map recorded on sample S2 in the suspended and substrate-supported regions, with three representative positions, highlighted by red filled circles and labeled A, B, and C, based on their fluorescence intensity.
  (c, d, e) Zero-field Optically Detected Magnetic Resonance (ODMR) spectra for the three representative positions shown in (b). The dots are the experimental data and the solid lines are the double-Gaussian fit.
  (f) Plot of the zero-field splitting (ZFS) of the ODMR-spectral dip as a function of fluorescence counts for various positions in the hBN sample, extracted from the ODMR scans in (c, d, e). The '$\times$' notation indicates the fluorescence increase relative to the dark spot on the hBN.}
  \label{fig3}

\end{figure}


Since the possibility that the bright spot in the suspended region contains a larger number of ${V_{\text{B}}^-}$ centers than the dark spot has been ruled out, the enhanced fluorescence comes from an intrinsic enhancement of the fluorescence emission of each ${V_{\text{B}}^-}$ center. The suspended regions studied so far occur by accident during the transfer process in flakes that exhibit one or more wrinkles. A wrinkle is a complex entity involving not only a lot of strain but also structural defects that typically will be correlated with the strain deformations and local symmetry breaking. 
In order to investigate whether the fluorescence enhancement originates from the difference in local structural deformation induced during the ion implantation, we create suspended regions in a deterministic fashion by selecting hBN flakes that are composed of regions of different thicknesses. If a thinner region of hBN is surrounded by thicker portions, a suspended area is obtained by placing the flake upside down on the substrate (the gold stripline) as depicted in Fig. \ref{fig3}a. This approach ensures a minimal difference between suspended and supported regions before He$^{+}$-implantation.
Then He$^{+}$-implantation with the same dose as for wrinkled devices completes the fabrication. 
This procedure minimizes the overall strain and highlights the impact of a suspended or supported region on the ion implantation process. 

\begin{figure}[h!]
  \centering
  \includegraphics[width = \textwidth]{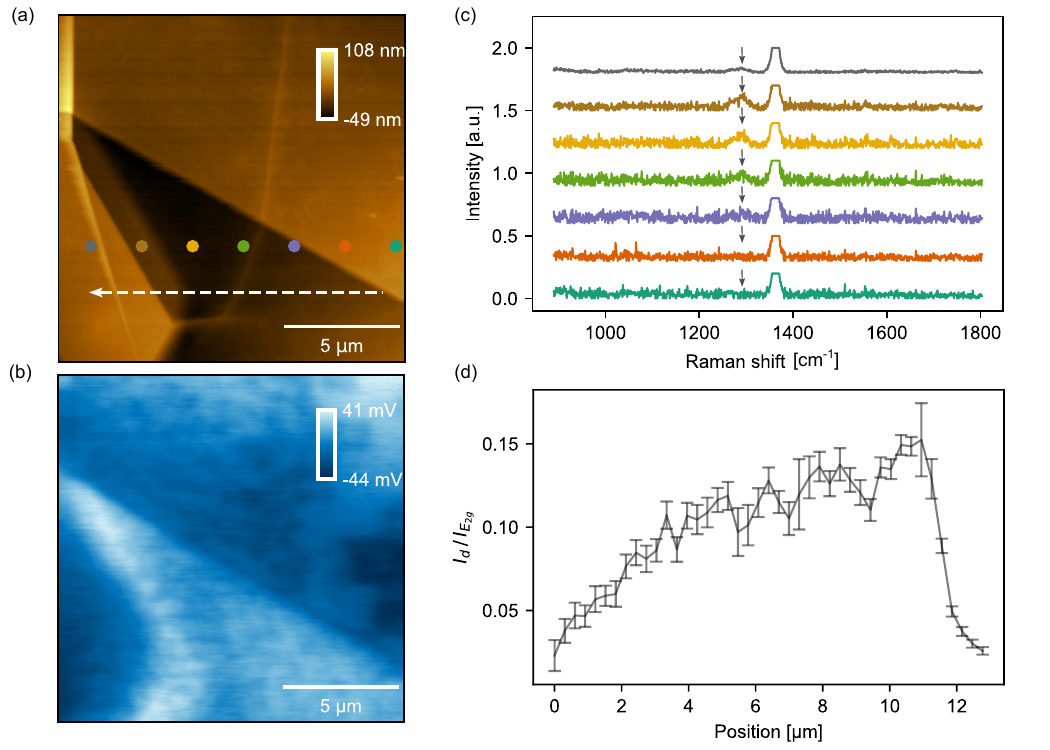}
  \caption{(a) AFM scan of hBN sample S2, showing regions of varying thickness. Several representative positions are labeled in different colors.
  (b) KPFM scan of the hBN sample S2 over a similar region, revealing surface potential variations at different positions.
  (c) Raman spectra at the representative positions corresponding to the color labels. Curves are offset with 0.3 on the vertical axis for the sake of clarity. A higher intensity of the $E^{'}$ peak at $\sim$1290 cm$^{-1}$ indicates increased deformation.
  (d) Linecut of the ratio between the intensity of the $E^{'}$ peak and the intensity of the hBN $E_{\text{2g}}$ peak, as indicated by the white line in (a).}
  \label{fig4}
\end{figure}

An atomic force microscopy (AFM) scan is performed on the device to confirm that the hBN flake indeed contains thinner suspended and substrate-supported regions (Sample S2, Fig. \ref{fig4}a).
The fluorescence emission from the thinner suspended hBN region is significantly larger than that from hBN areas that are supported by the substrate (Fig. 3b). The enhanced fluorescence confirms that there is a difference between suspended and supported regions during the ion implantation. Since SRIM simulations have shown that the created vacancy concentration in the suspended region is comparable to that in supported regions (Supplemental Section 2), the different behavior is attributed to local deformations caused by the collisions of the He$^+$-ions with the hBN lattice. On the one hand, different deformations for suspended and supported regions should be expected because of the different mechanical flexibility; On the other hand, local deformation of the hBN lattice could enhance the fluorescence emission by reducing the symmetry as discussed before.

Lattice deformations are always accompanied by local strain and field, which shifts and splits the spin sublevels of ${V_{\text{B}}^-}$ centers. At zero magnetic field, the ground-state energy levels of ${V_{\text{B}}^-}$ centers are dependent on the zero-field splitting (ZFS) parameters $D_\text{gs}$ and $E_\text{gs}$, as depicted in Fig. \ref{fig1}b. The separation between the $m_s=+1$ and $m_s=-1$ sublevels, $2E_\text{gs}$, is proportional to the transverse strain. Zero-field ODMR could be utilized to reveal the strain-induced separation. This is illustrated in Fig. \ref{fig3}c, d and e, where we have measured the zero-field ODMR spectra for three different locations (marked as A, B and C in panel b) with different fluorescence counts. Each spectrum shows two dips at the frequencies $\nu_\pm=(D_\text{gs}\pm E_\text{gs})/h$, corresponding to the respective transition between $m_s=0$ and $m_s=\pm1$. Here, $h$ is the Planck constant. The ZFS of the spectral dip, $\nu_+-\nu_-=2E_\text{gs}/h$, shows a positive correlation with the fluorescence count. As shown in Fig. \ref{fig3}f, the strain-induced ZFS is more than 10 MHz larger in the suspended region (location A) compared to the region where the hBN is supported by the substrate (location C). The fluorescence count is about 10 times larger as well. Similar behavior has also been observed in suspended regions on other samples (see Section 5, SI). The strain alone does not have such a significant impact on the fluorescence of ${V_{\text{B}}^-}$ centers as shown in a recent study\cite{yangSpinDefectsHexagonal2022}.
These results suggest that the ZFS reflects the deformation potential, which is accompanied by local strain, fields, and other perturbations that influence fluorescence. Implantation with the same dose results in a larger deformation in suspended regions than in the hBN areas supported by the substrate.


The deformation of the hBN lattice also modulates the electronic structure and work function of the hBN flake. Different local deformations in the suspended and supported regions lead to variations in the surface potential. This can be visualized by conducting KPFM measurements. The KPFM technique has been used to image nanoscale surface potential variations caused by the Moiré-induced ferroelectric effect in twisted hBN\cite{woods2021charge,vizner2021interfacial}. Here, we perform a KPFM scan over the suspended and supported hBN regions to measure the spatial distribution of the surface potential. The measurement result is shown in Fig. \ref{fig4}b. Indeed, we observe a significant difference in the surface potential between the suspended and supported regions. 
This is consistent with the differences in the local deformation as observed from the fluorescence count (Fig. \ref{fig3}b).
Similar spatial distributions of the surface potential have been observed on other hBN samples after implantation (see Section 5, SI), indicating that in general suspended regions suffer larger deformations compared to substrate-supported areas. KPFM measurements on hBN samples prior to implantation show no significant difference in surface potential between suspended and supported regions (see Section 6, SI), confirming that the deformations are induced by the ion implantation. 

Moreover, the charge state of $V_\text{B}^-$ centers can be influenced by local deformations as these modify the electronic structure. Since the photoluminescence spectra of boron vacancy ($V_\text{B}$) centers in other charge states (such as the neutral $V_\text{B}^0$ and doubly negatively charged $V_\text{B}^{2-}$) are unknown, the charge-state transformation of $V_\text{B}$ centers is generally characterized by a change in the photoluminescence spectrum or fluorescence intensity of the $V_\text{B}^-$ centers\cite{gale2023manipulating, NanoLett2025ChargeTuning}. 
An increase in the $V_\text{B}^-$ spectral or fluorescence intensity indicates a transformation of $V_\text{B}$ centers from other charge states into $V_\text{B}^-$, and vice versa.
Following the method which has been used to selectively measure the charge-state dynamics of NV centers in diamond\cite{PRB2019ChargeNV}, we apply a strong off-axis magnetic field to quench the spin polarization and measure the spin-independent dynamics of $V_\text{B}$ centers in the dark (see Section 4, SI). 
A laser pulse is applied to detect the fluorescence from $V_\text{B}^-$ centers after the dark time. The detection window is short so that the fluorescence intensity is proportional to the number of $V_\text{B}^-$ centers before detection.
An increase in the fluorescence intensity is observed with increasing dark time for the $V_\text{B}$ centers in the suspended region, while a negligible change in the fluorescence intensity is observed for the $V_\text{B}$ centers in areas supported by the substrate. The different behavior of the $V_\text{B}$ charge-state dynamics indicates different local deformations in the suspended and supported regions.

Apart from the electronic structure, also the phonon band structure can be modified by the deformation of the hBN lattice. To investigate the phonon band structure, we perform confocal Raman spectroscopy on the hBN flake starting from the supported region to the suspended region. As shown in Fig. \ref{fig4}c, the Raman spectra measured in the suspended region show an additional peak at $\sim$1290 ${\text{cm}}^{-1}$, which can hardly be observed in the spectra measured in the supported region. This additional peak, denoted as the $E^{'}$ peak in the following, was believed to be associated with LO phonon mode of cubic boron nitride (cBN) according to previous reports\cite{machaka2010formation}.
In recent years, this scenario has been questioned and excluded, because the TO phonon mode of cBN is absent in the Raman spectra\cite{sarkar2023identifying}. 
Ref. \cite{venturiSelectiveGenerationLuminescent2024a} now attributes the $E^{'}$ peak to a vibrational mode of the $V_\text{B}^-$ centers. 
While the frequent positive correlation between the intensity of the Raman $E^{'}$ peak and the fluorescence intensity of the $V_\text{B}^-$ centers lends support to this interpretation, so far there is no evidence from experiments that offer additional variables that can be well controlled.
Moreover, there are also instances where the $E^{'}$ peak has a stronger intensity whereas the fluorescence intensity of the $V_\text{B}^-$ centers is lower.
In particular, the Raman $E^{'}$ peak has not been observed for $V_\text{B}^-$ centers created by electron irradiation\cite{Nanomater2021VBElectIrrad}. While an analysis of the microscopic origin of the Raman $E^{'}$ peak is beyond the scope of this manuscript, we believe that the $E^{'}$ peak is related to the lattice deformation since it is the only possible significant difference between suspended and supported regions in our experiments. For example, the $E^{'}$ peak could be a vibrational mode of the $V_\text{B}^-$ centers activated by the lattice deformation induced symmetry reduction.This would explain the complexity of the correlation between the $E^{'}$ peak intensity and the $V_\text{B}^-$ fluorescence intensity, since both of them are dependent on multiple variables such as the lattice deformation and the number of $V_\text{B}^-$ centers. 
This proposition that the $E^{'}$ mode can be activated upon reduction of the crystal symmetry has been put forward previously in Ref. \cite{sarkar2023identifying} and is also consistent with experimental observations in MoS$_2$ that is irradiated either with Ar$^+$ ions or protons\cite{PRAppl2017VibraModeMoS2, APL2012MagnetMoS2}.

We quantitatively characterize the local deformation in the suspended and supported regions by determining the ratio of the intensity of the Raman $E^{'}$ peak ($I_\text{d}$) and the intensity of the intrinsic hBN $E_\text{2g}$ phonon peak ($I_{E_\text{2g}}$) at $\sim$1365 ${\text{cm}}^{-1}$. As shown in Fig. \ref{fig4}d, the intensity ratio $I_\text{d}/I_{E_\text{2g}}$ increases when crossing the boundary from the supported to the suspended region. It reaches a maximum in the suspended region and then decreases again upon crossing the other boundary to the supported region. 
Hence, the Raman peak intensity exhibits a spatial distribution similar to that of the deformation potential. It is larger in the suspended region compared to the supported regions.
Due to the diffraction limit, each Raman spectrum represents the signal integrated over a focal volume of about 500 nm in diameter. Therefore, the intensity ratio $I_\text{d}/I_{E_\text{2g}}$ characterizes the average deformation, while the linewidth of the $E^{'}$ peak reflects the inhomogeneity of the deformation potential within the focal volume. Considering the random nature of the deformation potential caused by the collisions of the He$^+$ ions with the hBN lattice, the average and the spatial inhomogeneity of the deformation potential should be positively correlated. 
Indeed, as shown in Section 7 of the SI, spatial distribution of the linewidth of the $E^{'}$ peak is similar to that of the intensity ratio $I_\text{d}/I_{E_\text{2g}}$.

The measurements of the zero-field ODMR, KPFM, and Raman spectroscopy all identify differences in the deformation potential between the suspended and the substrate-supported regions of the hBN flake. 
The collision with He$^+$ ions can cause larger deformations in the suspended regions due to the larger mechanical flexibility.
The deformation breaks the D$_\text{3h}$ symmetry of the ${V_\text{B}^-}$ centers.
The optical transition from the excited state $^3E''$ to the ground state $^3A_2'$ of the ${V_\text{B}^-}$ centers is no longer forbidden to first order by symmetry and, as a result, the fluorescence emission is enhanced.
The enhanced fluorescence intensity of the ${V_\text{B}^-}$ centers in the suspended region is positively correlated with larger local deformations as a result of symmetry reduction.


\section{Conclusions}
In summary, we have developed an effective approach to enhance the fluorescence emission of ${V_\text{B}^-}$ centers by simply targeting and fabricating suspended hBN regions. 
This increases the emission of ${V_\text{B}^-}$ centers by more than a factor of 30, while preserving all its characteristic spin properties, including the ODMR contrast, the linewidth, the $T_1$ time, etc.
The enhancement of ${V_\text{B}^-}$ emission is attributed to implantation-induced deformation in the suspended hBN region. This is confirmed by the measurements of the zero-field ODMR splitting, confocal Raman spectroscopy, KPFM, and charge-state dynamics of the ${V_\text{B}^-}$ centers. Our work highlights the critical role of symmetry in the fluorescence emission of ${V_\text{B}^-}$ centers and demonstrates that the deformation of the hBN lattice can activate optical transitions that are otherwise forbidden by symmetry. 
These local deformations can also promote the formation of other optically active spin defects in hBN, such as carbon-related defect centers, and enhance their emission. In our previous work \cite{chejanovsky2021single}, the majority of optically addressable spin defects were found near wrinkles, suggesting that controlled local deformations are a promising route to tailor the optical and spin properties of hBN defects. More generally, our results offer new insights into engineering the quantum properties of 2D materials by manipulating structural deformations and defect symmetry. This is relevant not only to hBN but also to other hosts such as transition metal dichalcogenides\cite{NC2022_AntisiteDefectQubitTMD, NC2022_CarbonQubitWS2} and even SiC and diamond. The approach expands the possibilities for quantum sensing technology.

\begin{acknowledgement}
We acknowledge the financial support from the DFG via FOR 2724, the BMFTR via Cluster4Future QSENS, and the EU via project Amadeus and C-QuEnS. J.G. acknowledges the financial support from the National Natural Science Foundation of China (Grant No. 12504563).

\end{acknowledgement}
\bibliography{achemso-demo}

\end{document}